\documentclass[11pt]{article}
\usepackage{amsmath}
\usepackage[preprint]{acl}

\usepackage{times}
\usepackage{latexsym}
\usepackage{amssymb}
\usepackage{booktabs} %
\usepackage{siunitx}  %
\usepackage{tabularx}
\usepackage{xurl} 
\usepackage[T1]{fontenc}

\usepackage[utf8]{inputenc}

\usepackage{microtype}

\usepackage{inconsolata}

\usepackage{graphicx}

\title{RvB: Automating AI System Hardening via Iterative Red-Blue Games}

\author{
    \textbf{Lige Huang\textsuperscript{1,2}\thanks{\ \ Equal contribution.}} \quad
    \textbf{Zicheng Liu\textsuperscript{1,3}\footnotemark[1]} \quad
    \textbf{Jie Zhang\textsuperscript{1}\footnotemark[1]} \\
    \textbf{Lewen Yan\textsuperscript{1}} \quad
    \textbf{Dongrui Liu\textsuperscript{1}} \quad
    \textbf{Jing Shao\textsuperscript{1}\thanks{\ \ Corresponding author.}} \\
    \\
    \textsuperscript{1}Shanghai Artificial Intelligence Laboratory \quad \\
    \textsuperscript{2}Institute of Information Engineering, Chinese Academy of Sciences \\
    \textsuperscript{3}Shanghai Jiao Tong University \\
    \\
    \texttt{ryukosei@sjtu.edu.cn} \\
    \texttt{\{huanglige, zhangjie1, shaojing\}@pjlab.org.cn} \\
}

\begin{document}
\maketitle
\begin{abstract}
The dual offensive and defensive utility of Large Language Models (LLMs) highlights a critical gap in AI security: the lack of unified frameworks for dynamic, iterative adversarial adaptation hardening. To bridge this gap, we propose the Red Team vs. Blue Team (RvB) framework, formulated as a training-free, sequential, imperfect-information game. In this process, the Red Team exposes vulnerabilities, driving the Blue Team to learning effective solutions without parameter updates. We validate our framework across two challenging domains: dynamic code hardening against CVEs and guardrail optimization against jailbreaks. Our empirical results show that this interaction compels the Blue Team to learn fundamental defensive principles, leading to robust remediations that are not merely overfitted to specific exploits. RvB achieves Defense Success Rates of 90\% and 45\% across the respective tasks while maintaining near 0\% False Positive Rates, significantly surpassing baselines. This work establishes the iterative adversarial interaction framework as a practical paradigm that automates the continuous hardening of AI systems.
\end{abstract}

\section{Introduction}

The rapid proliferation of Large Language Models (LLMs) has introduced a fundamental duality to the field of security \citep{li2025security}. On one hand, LLMs serve as powerful offensive instruments, exemplified by autonomous agents like \textit{Pentest-Agent} \citep{shen2025pentestagentincorporatingllmagents} and collaborative frameworks like \textit{CAI} \citep{mayoralvilches2025caiopenbugbountyready}, which can independently execute complex exploits, as well as orchestrate sophisticated jailbreaking attacks \citep{mazeika2024harmbench}. On the other hand, they act as sophisticated defensive shields, enabling systems such as \textit{SWE-agent} \citep{yang2024sweagent} and defensive adapters \citep{Castro_2025} to automate vulnerability patching and incident response, and function as dynamic guardrails to prevent harmful content generation \citep{zeng2024autodefense}.

However, despite these parallel advancements, offensive and defensive research streams remain largely disjoint \citep{zou2025security}. Defensive frameworks predominantly rely on static benchmarks or post-hoc analysis, limiting their capacity to anticipate novel attack vectors \citep{chao2024jailbreakbenchopenrobustnessbenchmark}. Conversely, offensive agents often operate without a responsive adversary, failing to assess the target's resilience under worst-case adversarial scenarios. This isolation highlights a critical gap: the absence of a unified game-theoretic paradigm that forces these systems to engage in a continuous cycle of iterative adversarial adaptation, thereby proactively discovering and mitigating unknown vulnerabilities \citep{huynh2025understanding}.

\begin{figure}[!t]
    \centering
    \includegraphics[width=\linewidth]{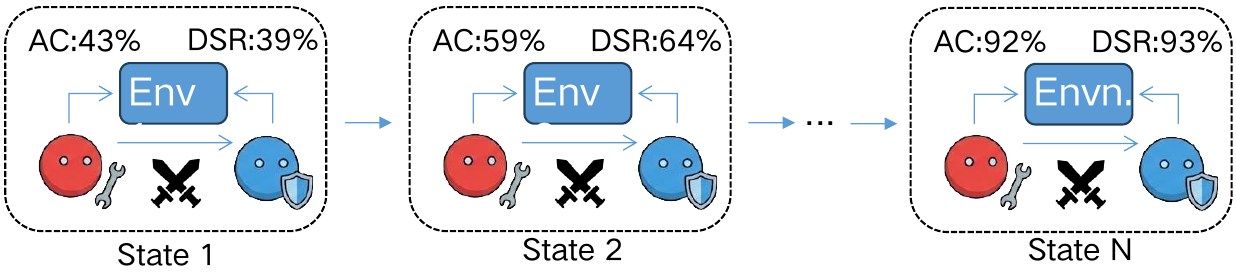}
    \caption{The Red vs. Blue (RvB) framework across successive states ($S_1 \rightarrow S_N$). 
    The \textbf{Attack Complexity (AC)} represents the escalating sophistication and specificity of exploits required to bypass the hardening environment as epistemic uncertainty decreases. 
    The \textbf{Defense Success Rate (DSR)} measures the Blue Team's growing efficacy in neutralizing vulnerabilities through iterative fixes. }
    \label{fig:rvb_evolution}
\end{figure}

To bridge this gap, we propose a novel \textbf{Red Team vs. Blue Team (RvB)} framework. In contrast to traditional reinforcement learning paradigms that are constrained by high sample complexity and necessitate expensive model fine-tuning, our framework is formulated as a training-free, sequential, imperfect-information game. As illustrated in Figure~\ref{fig:rvb_evolution}, we model security hardening as an iterative turn-based interaction. In each round, the Red Team (Attacker) interacts with the dynamic environment to execute multi-step exploits. Conversely, the Blue Team (Defender) uses attack logs as a basis but conducts autonomous interrogation, probing the system state to pinpoint the root cause and synthesize a fix. This information asymmetry forces the defender to transcend superficial log pattern matching; instead, it must employ inductive reasoning to infer the underlying vulnerability logic, thereby generating generalized patches that secure the system against potential variants.

This paradigm shifts the focus from static benchmarking to dynamic adversarial hardening. By capitalizing on the in-context reasoning capabilities of foundation models within a persistent environment, our framework establishes a virtuous cycle where the adversarial interplay directly translates into system robustness. Crucially, rather than permanently enhancing the agents themselves, this process iteratively adapts the \textit{security artifacts}: the Red Team exposes latent, high-complexity vulnerability paths, necessitating the Blue Team to construct more comprehensive and resilient defense protocols. This ensures that the target system is hardened against not merely specific exploits, but broader classes of emergent threats.

We validate the versatility and effectiveness of our RvB framework across two distinct and challenging domains:
\begin{itemize}
    \item \textbf{Cyber Security (Code Hardening):} We deploy agents in a realistic network environment to autonomously identify and patch logic vulnerabilities (e.g., CVEs). The Red Team utilizes tool-use capabilities to penetrate the system, forcing the Blue Team to analyze execution logs and generate functional code patches that pass rigorous regression tests.
    \item \textbf{Content Security (Guardrail Optimization):} We simulate an adversarial dialogue between a Jailbreaker and a Guardrail. The Red Team iteratively optimizes prompt injection strategies, driving the Blue Team to dynamically refine its safety rules to block diverse jailbreak attempts effectively.
\end{itemize}

Our empirical results demonstrate that this iterative adversarial adaptation significantly outperforms baselines. Across dynamic code hardening and guardrail optimization tasks, RvB achieves Defense Success Rates of 90\% and 45\%, respectively, while maintaining near-zero False Positive Rates. Crucially, the system exhibits robust generalization to unseen attacks. Furthermore, it maintains a computational cost comparable to cooperative baselines, even reducing token consumption by over 18\% in code remediation scenarios. Collectively, these findings establish RvB as a practical, training-free paradigm that automates the continuous hardening of AI systems.

\section{Related Work}

\subsection{Multi-Agent Systems (MAS)}

Multi-Agent Systems (MAS) have emerged as a powerful paradigm for solving complex problems that exceed the capabilities of a single agent, fostering emergent collective intelligence through interaction, coordination, and collaboration \citep{dorri2018multi}. With the advent of LLMs, a new wave of research has focused on building LLM-based agents capable of sophisticated, knowledge-intensive behaviors. A prominent line of this work centers on \textbf{cooperative problem-solving}, where agents with a shared goal employ diverse strategies, such as decentralized coordination or collaborative reasoning \citep{tao2024magis, wang2024multi, ishibashi2024self, yang2025agentnet, jha2025cross}. 

However, the assumption of a fully trusted context in these cooperative frameworks limits their applicability in security-critical domains. To address this, another significant branch of MAS research explores \textbf{adversarial interactions}. Studies have developed agents capable of executing sophisticated attacks, such as prompt injection, communication attacks, or exploiting system weaknesses through adversarial prompting \citep{lee2024prompt, he2025red, arora2025exposing}. Concurrently, defensive frameworks have emerged, utilizing agents to defend against specific threats like jailbreaks or backdoor attacks \citep{zeng2024autodefense, fan2025peerguard}. While this adversarial dynamic drives capability enhancement, many existing works are either modeled as symmetric self-play or focus on specific, predefined attack-defense vectors. This often leaves a gap in addressing unknown threats and ensuring that the global objective remains a non-zero-sum goal of system hardening, rather than a zero-sum win for either the attacker or the defender.

\subsection{AI Security}

The application of LLMs to cybersecurity has introduced a fundamental duality between automated vulnerability discovery and remediation. On the offensive side, research focuses on automating penetration testing and fuzzing. Agents such as PentestGPT \citep{deng2024pentestgptllmempoweredautomaticpenetration} and AutoPentest \citep{henke2025autopentestenhancingvulnerabilitymanagement} leverage LLMs to decompose complex attack procedures, autonomously guiding standard security tools to exploit target systems. LLM-enhanced fuzzing frameworks, exemplified by TitanFuzz \citep{deng2023largelanguagemodelszeroshot} and Fuzz4All \citep{Xia_2024}, utilize LLMs to synthesize edge-case inputs that trigger deep-logic vulnerabilities in software libraries. Conversely, defensive research has pivoted toward Automated Program Repair (APR). Systems such as InferFix\citep{jin2023inferfixendtoendprogramrepair} and VulRepair\citep{de_Fitero_Dominguez_2024} employ fine-tuned models to localize and patch security defects based on static analysis. More recently, autonomous software engineering agents, such as SWE-agent\citep{yang2024sweagentagentcomputerinterfacesenable}, have demonstrated the capacity to resolve GitHub issues and apply functional patches in real-world repositories. However, these offensive and defensive streams operate largely in isolation: offensive agents lack a responsive patching mechanism, while defensive repair tools typically rely on static benchmarks rather than dynamic, adversarial feedback.

Within AI Safety, the adversarial adaptation between offensive Red Team and defensive Blue Team agents has become a critical research area for addressing jailbreak attacks. On the offensive side, autonomous agents systematically discover model vulnerabilities through diverse strategies. These include frameworks that compose human-defined principles \citep{xiong2025cop}, reframe harmful queries as reasoning tasks for self-jailbreaking \citep{ying2025reasoning}, evolve a library of reusable attack strategies \citep{liu2025automated}, or simulate multi-turn human attacks via multi-agent collaboration \citep{rahman2025x}. In response, adaptive Blue Team agents have emerged with a range of defensive approaches. Some operate at the prompt level by optimizing a universal defensive suffix \citep{zhou2024robustpromptoptimizationdefending, xiong2025defensivepromptpatchrobust}. Others function as external, training-free guardians, using either dynamic, user-defined policies \citep{hoover2025dynaguarddynamicguardianmodel} or a retrieval-augmented database of known attacks \citep{yang2025retrievalaugmenteddefenseadaptivecontrollable}. A third approach re-trains models to incorporate an internal, step-by-step safety reasoning process \citep{zhu2025reasoningtodefendsafetyawarereasoningdefend, yang2025enhancingmodeldefensejailbreaks}. This ongoing arms race between increasingly sophisticated red and blue agents is a primary driver of progress in building more robust and aligned LLMs.

\section{The RvB Framework: A Formal Paradigm}

\begin{figure*}[!t] 
    \centering
    \includegraphics[width=1.0\linewidth]{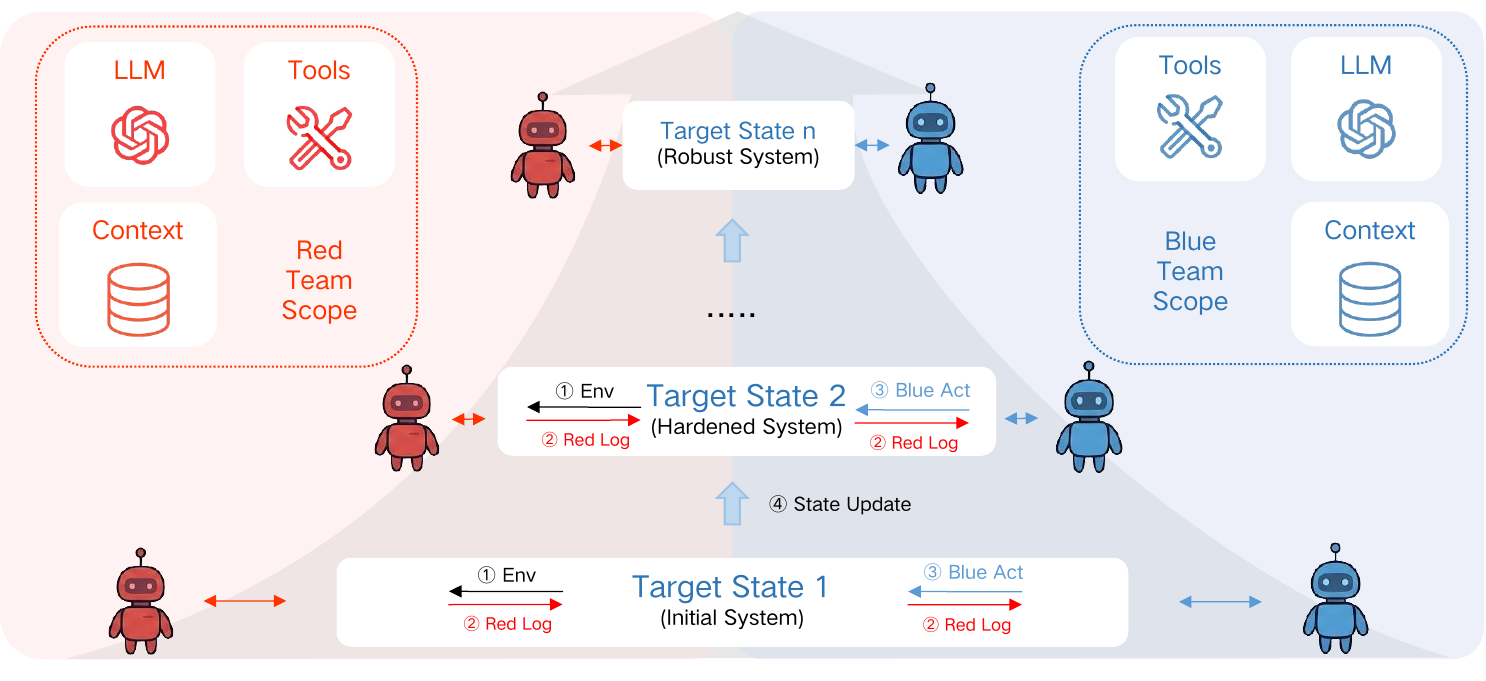}
    \caption{Overview of the RvB Framework. Modeled as a sequential, imperfect-information game, the process drives system hardening ($S_1 \to S_N$) through iterative adversarial interactions. The environment acts as an \textit{externalized memory}, necessitating the \textbf{iterative adversarial adaptation} of agent strategies without requiring internal model parameter updates.}
    \label{fig:framework_overview}
\end{figure*}

In this section, we formalize the \textbf{Red Team vs. Blue Team (RvB)} framework as a training-free, sequential, imperfect-information game. In contrast to traditional Multi-Agent Reinforcement Learning (MARL) which relies on internal policy updates, our paradigm leverages \textit{Externalized Memory} (Figure~\ref{fig:framework_overview}). Although the agents remain computationally stateless and reset between rounds, the theoretical Bayesian belief updates are \textit{manifested} through the persistent state transitions of the target system.

\subsection{Formal Model Setup}
\label{sec:eq}
We define the game dynamics via the following tuple:

\paragraph{Sequential Interaction.} The game proceeds in discrete rounds $k = 1, 2, \dots, N$, progressing from an initial vulnerable state to a robust state. Let $\Pi_R$ and $\Pi_B$ denote the fixed meta-strategy spaces (i.e., agent capabilities and toolsets) for the Red and Blue teams, respectively. In our training-free setting, $\Pi$ remains static, no gradient updates occur.

\paragraph{State as Externalized History.} Let $M_k$ denote the interaction log history up to round $k$. We define the \textit{Target System State} $S_k$ as a physical encoding of this history, i.e., $S_k \cong \text{Encode}(M_k)$. The transition $S_k \to S_{k+1}$ represents the consolidation of historical interaction data into the environment.

\paragraph{Belief Distribution.} Operating under partial observability, the Red Team observes the state $S_k$ but not the Blue Team's specific defense logic $\pi_B$. The Red Team's \textit{Belief Distribution(b)} at the start of round $k$ is conditioned directly on the observable state:
\begin{equation}
    b_k(\pi_B) = P(\pi_B \mid M_k) \equiv P(\pi_B \mid S_k).
\end{equation}

\subsection{Mechanism: Belief Update via Environmental Feedback}

Capability enhancement is modeled not as parameter tuning, but as the refinement of the belief distribution $b_k$ through the interaction loop.

\paragraph{Decision Process (Subjective Expected Utility).}
At round $k$, the Red Team selects an action $a_R^k$ to maximize utility based on the current belief $b_k$. The agent computes the Subjective Expected Utility (SEU):
\begin{equation}
    a_R^k = \arg\max_{a_r \in \mathcal{A}_R} \mathbb{E}_{\pi_B \sim b_k} [ U_R(a_r, \pi_B) ],
\end{equation}
where $U_R$ is the reward function. In early rounds ($S_1$), the high-entropy belief results in generalized exploration.

\paragraph{Posterior Update via State Transition.}
The execution logs $L_R$ generated by the Red Team trigger a remediation step by the Blue Team, inducing a state transition to $S_{k+1}$. This transition acts as the Evidence ($E_k$). The agent updates its belief to the posterior $b_{k+1}$ using Bayes' Theorem. 
\begin{equation}
\label{eq:bayes_update}
\begin{aligned}
    b_{k+1}(\pi_B) &= P(\pi_B \mid E_k, S_k) \\
    &= \frac{P(E_k \mid \pi_B, S_k) \cdot b_k(\pi_B)}{\sum_{\pi' \in \Pi_B} P(E_k \mid \pi', S_k) \cdot b_k(\pi')}.
\end{aligned}
\end{equation}
\noindent\textbf{The Stateless Implementation.} Since the agent is reset, this posterior calculation is realized by the \textit{filtering mechanism} of the environment. The new state $S_{k+1}$ physically invalidates the subset of strategies exploitable in $S_k$. Consequently, when the Red Team observes $S_{k+1}$, its effective search space is mathematically equivalent to sampling from the posterior $b_{k+1}$.

\subsection{Information-Theoretic Analysis}

We quantify capability enhancement as the reduction of epistemic uncertainty regarding the optimal attack vector. Since the state transition provides information gain (assuming non-negative mutual information), the entropy of the agent's belief monotonically decreases:
\begin{equation}
\label{eq:belief}
    H(b_{k+1}) \le H(b_k).
\end{equation}
This inequality formalizes the system's progression:
\begin{itemize}
    \item \textbf{Exploration ($k=1$):} Maximal entropy $H_1$ implies a uniform prior, leading to trial-and-error attacks.
    \item \textbf{Exploitation ($k\to N$):} As the environment accumulates defenses, the evidence $E_{1:k}$ prunes the feasible strategy space. The belief $b_N$ becomes low-entropy (sharp), compelling the agent to execute highly specific, sophisticated exploits to bypass the accumulated constraints.
\end{itemize}

\section{Experiment}

\subsection{The Attacker vs. Defender in Cyber Security}
\paragraph{Experiment Setting}

In this experiment, we selected 10 specific vulnerability, including CVE-2022-30887, within the \textit{Pharmacy Management System v1.0}. We constructed a vulnerable PHP service environment by deploying the application's source code within a Docker container. We employed \textbf{CAI} \citep{mayoralvilches2025caiopenbugbountyready} as the red team agent within our red-blue teaming paradigm. Equipped with specialized interfaces such as Bash and the Model Context Protocol (MCP), this agent is capable of performing attack prediction against the network environment, dynamically probing the system, and generating comprehensive vulnerability reports. For the blue team agent, we selected the \textbf{Mini-SWE-Agent} \citep{yang2024sweagent}, which functions by probing the system status and utilizing Bash commands to directly modify the source code, thereby achieving the remediation of environmental vulnerabilities.

\begin{figure}[t!]
    \centering
    \includegraphics[width=1.0\linewidth]{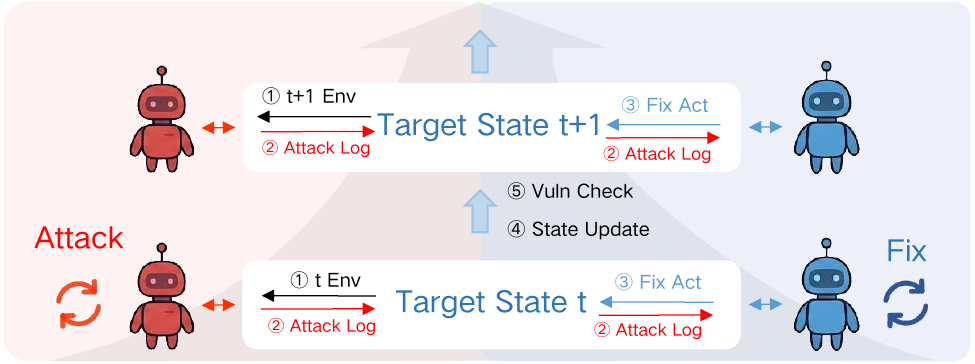}
    \caption{Overview of the iterative Red-Blue adversarial loop. At each state $t$, the Red Team probes the environment (1) to generate a vulnerability report (2). The Blue Team utilizes this report to apply a patch (3), updating the system to state $t+1$ (4). A final verification (5) confirms if the vulnerability is mitigated.}
    \label{fig:Cyber_Exp_case}
\end{figure}

The entire process is structured as a multi-round game where the two teams drive the iterative adversarial adaptation of offensive and defensive capabilities. We established a closed-loop adversarial workflow to simulate the continuous evolution of system security, as illustrated in Figure \ref{fig:Cyber_Exp_case}. A detailed breakdown of the agent architectures, model selections, and the step-by-step procedure is provided in Appendix~\ref{sec:appendix_exp1}.

The optimization process terminates if any of the following \textbf{stopping criteria} are met: 
(i) the failure of both the Red and Blue teams to generate any new structured findings (\textbf{null production}); 
(ii) an \textbf{execution failure} within the batch remediation module; 
(iii) \textbf{metric convergence}, defined as the vulnerability count remaining unchanged for three consecutive epochs ($\texttt{Count\_delay} = 3$); or 
(iv) reaching the maximum number of \textbf{predefined epochs} ($\texttt{max\_epoch} = 5$). 
To facilitate quantitative evaluation, we archive all interaction artifacts from each epoch and trace the vulnerability reduction trajectory, denoted as $(C_{i-1}, C_i)$.

\paragraph{Metrics}

To rigorously validate the effectiveness of the RvB framework, we established a metric suite centered on defensive robustness. We measure the per-round \textbf{Defense Success Rate (DSR)} to evaluate the Blue Team's remediation capabilities and employ the \textbf{Attack Success Count (ASC)} as a quantitative proxy for Attack Complexity (AC). Crucially, to guarantee fidelity in automated remediation, we proposed a granular decomposition of DSR into \textbf{True DSR (TDSR)} and \textbf{Fake DSR (FDSR)}. This distinction allows us to isolate effective semantic fixes from destructive patches that technically mitigate vulnerabilities only by inducing service outages or interface failures. The formal definitions and calculation logic for these metrics are detailed in Appendix~\ref{sec:metric_definitions}.

\paragraph{Results and Analysis}
Based on the experiments described above, we obtained the following key results:
\begin{figure}[t!]
    \centering
    \includegraphics[width=1.0\linewidth]{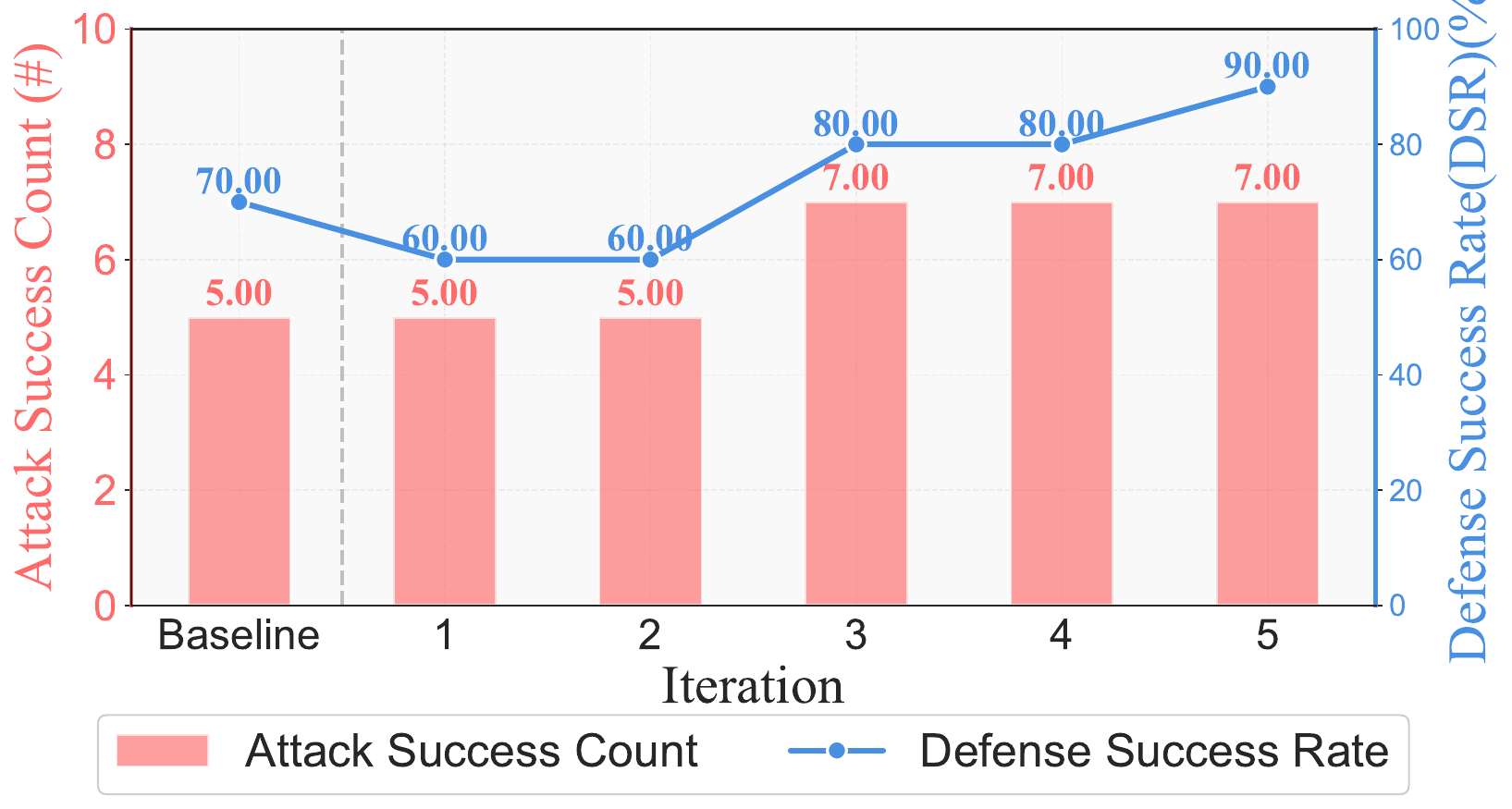} 
    \caption{Performance trajectory of the Red and Blue agents across 5 iterations. The bar chart (left axis) represents the cumulative count of successful exploits discovered by the Red Team, while the line graph (right axis) tracks the Blue Team's Defense Success Rate (DSR). The convergence toward a high DSR alongside sustained attack intensity validates the framework's effectiveness in automated security hardening.}
    \label{fig:Cyber_result_1}
\end{figure}

\textbf{1) The RvB framework significantly enhances cybersecurity operational capabilities.} 
As illustrated in Figure \ref{fig:Cyber_result_1}, the adversarial interaction effectively drives performance gains for both agents. We established a \textit{Baseline} simulating a standard cooperative Multi-Agent System (MAS), where the Defense Success Rate (DSR) and Attack Success Count (ASC) are calculated as the cumulative union of results over the maximum epochs. 
Observations indicate that while the RvB framework's initial DSR was lower than the baseline, it rapidly improved, surpassing the baseline at Iteration 3 (80\%) and reaching 90\% by Iteration 5. Concurrently, the Red Team demonstrated improved Attack Capability (AC) through state updates, identifying new vulnerabilities and increasing the ASC from 5 to 7. These results validate that the RvB paradigm effectively elevates the capabilities of both offensive and defensive agents compared to a static cooperative baseline.

\begin{figure}[t!]
    \centering
    \includegraphics[width=1.0\linewidth]{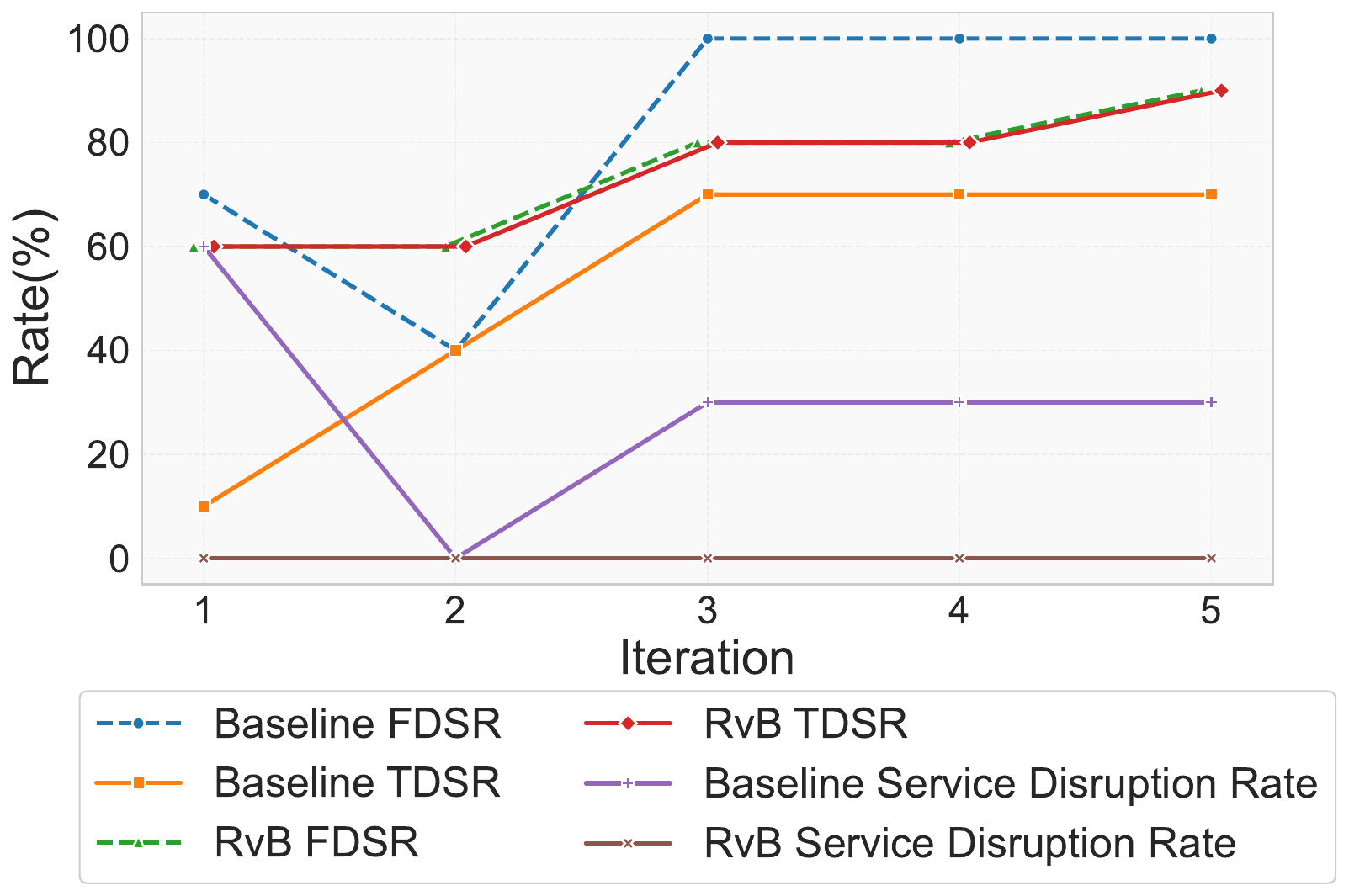} 
    \caption{Comparative analysis of defense robustness between the Baseline and RvB frameworks. The solid lines track the True Defense Success Rate (TDSR), representing valid fixes, while the dashed lines indicate the Fake Defense Success Rate (FDSR). The purple line highlights the significant Service Disruption Rate (SDR) in the Baseline, caused by destructive patches. In contrast, the RvB framework (brown line) maintains a near-zero SDR, demonstrating that adversarial feedback ensures high-quality remediation without compromising service availability.}
    \label{fig:Cyber_result_2}
\end{figure}

\textbf{2) The RvB framework improves the safety and robustness of the multi-agent system.} 
As shown in Figure \ref{fig:Cyber_result_2}, the system employing the RvB framework exhibits a substantially lower service drop rate. To quantify this, we defined three metrics: \textit{True DSR (TDSR)}, representing cases where the vulnerability is fixed and the service remains functional; \textit{Fake DSR (FDSR)}, where the attack check fails potentially due to service destruction (e.g., the Blue Team deleting critical files); and \textit{Service Disruption Rate (SDR)}, calculated as the difference between FDSR and TDSR.
Remarkably, the RvB framework maintains an SDR of 0 throughout the process. In contrast, the cooperative baseline exhibits an SDR as high as 60\%. Analysis of execution logs reveals that the Red Team's directed adversarial reports provide precise, actionable contexts, preventing the Blue Team from making indiscriminate or destructive modifications to the codebase. This confirms that the adversarial feedback loop is crucial for ensuring valid, non-disruptive remediation.

\subsection{The Jailbreaker vs. Guardrail in Content Security}

\paragraph{Experiment Setting}
To evaluate the RvB framework in a realistic security context, we designed an experiment simulating an adversarial dialogue between a \textit{Jailbreaker} (Red Team) and a \textit{Guardrail} (Blue Team). We selected harmful prompts from the HarmBench benchmark \citep{mazeika2024harmbench} as the attack targets. The Red Team agent, built upon the \textbf{Composition-of-Principles (CoP) framework} \citep{xiong2025cop}, iteratively refines attack strategies to generate effective jailbreak prompts. The Blue Team agent is an adaptive guardrail system based on \textbf{NeMo Guardrails} \citep{rebedea2023nemo}, which can dynamically augment its rule set to defend against novel attacks. 

\begin{figure}[t!]
    \centering
    \includegraphics[width=1.0\linewidth]{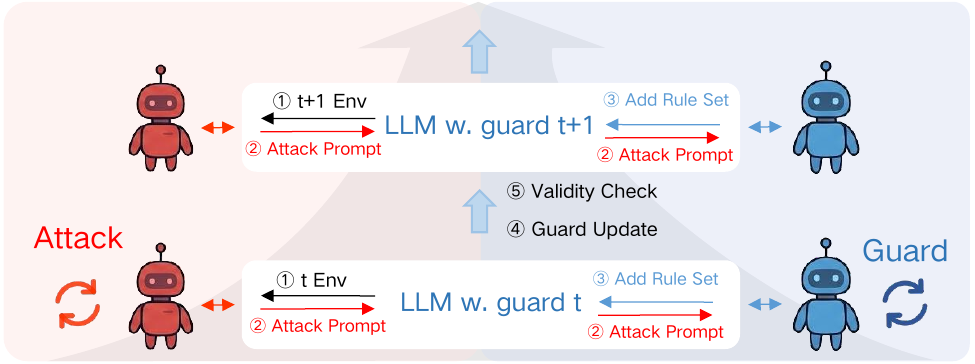}
    \caption{Overview of the iterative adversarial dialogue for guardrail optimization. The interaction cycle proceeds as follows: (1) The Red Team probes the current environment state; (2) it executes an adaptive attack prompt aimed at bypassing the guard; (3) The Blue Team analyzes the successful exploit to synthesize a targeted rule set; (4) The guardrail undergoes a dynamic update, transitioning the system to a more robust state $t+1$; and (5) A validity check using benign queries is performed to prevent over-interception and ensure the preservation of model utility.}
    \label{fig:Jailbreak_Exp_case}
\end{figure}

The entire process is structured as a multi-round game where the two teams drive the iterative adversarial adaptation of offensive and defensive capabilities, as illustrated in Figure~\ref{fig:Jailbreak_Exp_case}. A detailed breakdown of the agent architectures, model selections, and the step-by-step procedure is provided in Appendix \ref{sec:appendix_exp2_details}.

\paragraph{Metrics}
To quantitatively measure the dynamics of this adversarial interplay, we designed a set of corresponding metrics. To maintain a consistent defensive perspective, our primary metrics focus on the Blue Team's performance. We measure the \textbf{Defense Success Rate (DSR)} against the Red Team's latest attacks in each round, and the \textbf{Average Attack Turns (AAT)} the Red Team requires for a successful breach. For a more granular analysis of learning capabilities, we introduce the \textbf{Cross-Round Defense Efficacy (CRDE)} to test for catastrophic forgetting, and the \textbf{False Positive Rate (FPR)} to assess over-interception. The formal definitions of these metrics are provided in Appendix \ref{sec:appendix_metrics}.

\paragraph{Results and Analysis}
By compiling the metrics across four adversarial rounds, we obtained the following key results:

\begin{figure}[t!]
    \centering
    \includegraphics[width=1.0\linewidth]{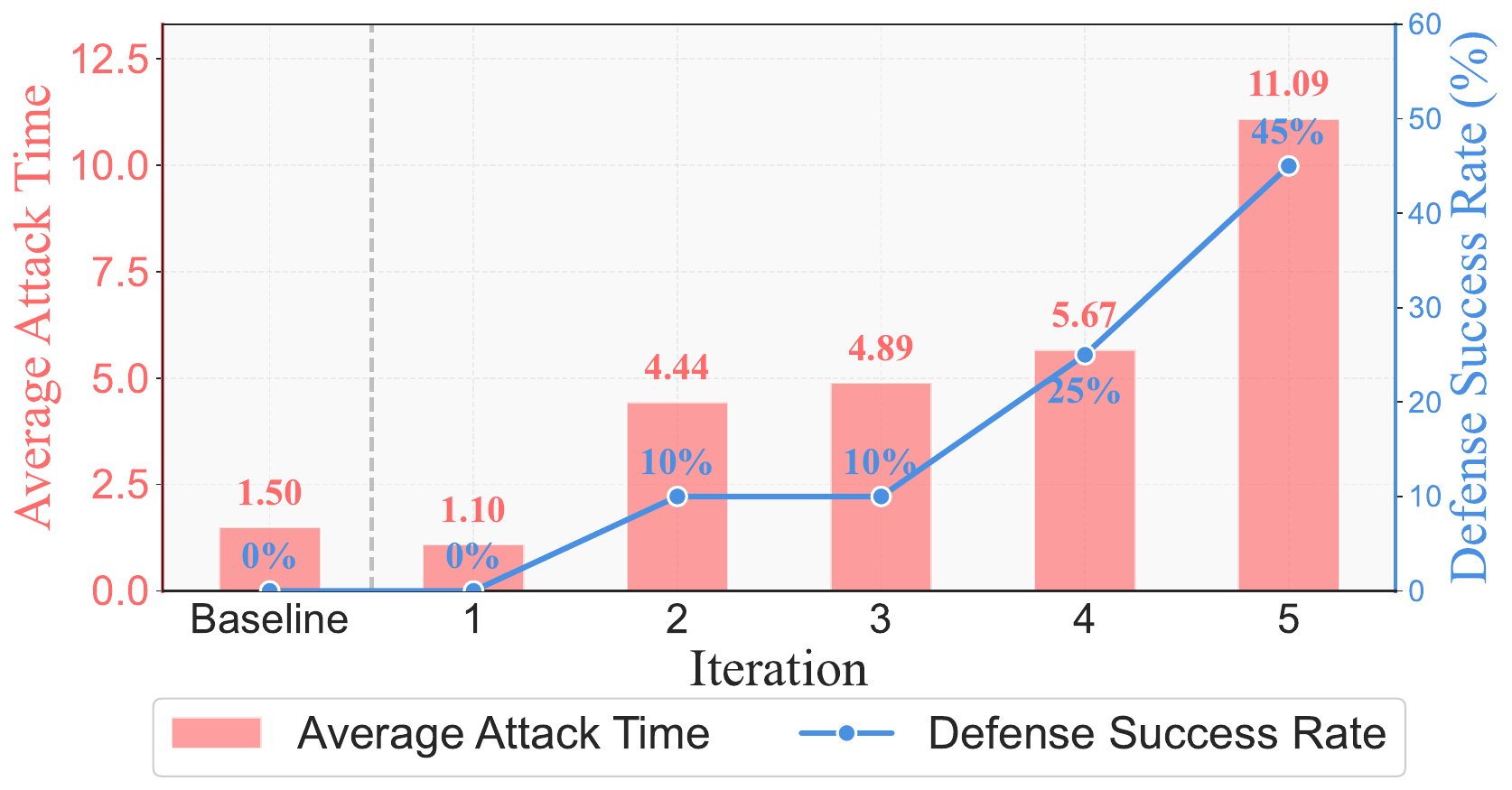} 
    \caption{Defense Success Rate and Average Attack Time comparison. As iterations progress, DSR (bars, right y-axis) consistently rises, while AAT (line, left y-axis) also trends upward, indicating that the Blue Team's guardrail is becoming progressively stronger and more difficult to bypass.}
    \label{fig:dsr_aat}
\end{figure}

\textbf{1) The RvB framework drives continuous and significant security improvement.} 
As depicted in Figure \ref{fig:dsr_aat}, the Blue Team's Defense Success Rate (DSR) against the Red Team's attacks shows a clear upward trend with each adversarial round, climbing from 0\% in the initial rounds to 45\% by the final iteration. This indicates that the guardrail's ability to intercept novel attacks is continuously improving. Concurrently, the Average Attack Turns (AAT) required for a successful breach also steadily rise, ultimately reaching 11.09 turns in the final round—a more than 7-fold increase from the baseline of 1.50. This further confirms that the Red Team's task is becoming progressively more difficult. The task-level analysis in Figure \ref{fig:jailbreak_heatmap} (in Appendix) corroborates this, showing that the number of attempts required for successful attacks increases substantially in later rounds for most tasks.

\begin{figure}[t!]
    \centering
    \includegraphics[width=1.0\linewidth]{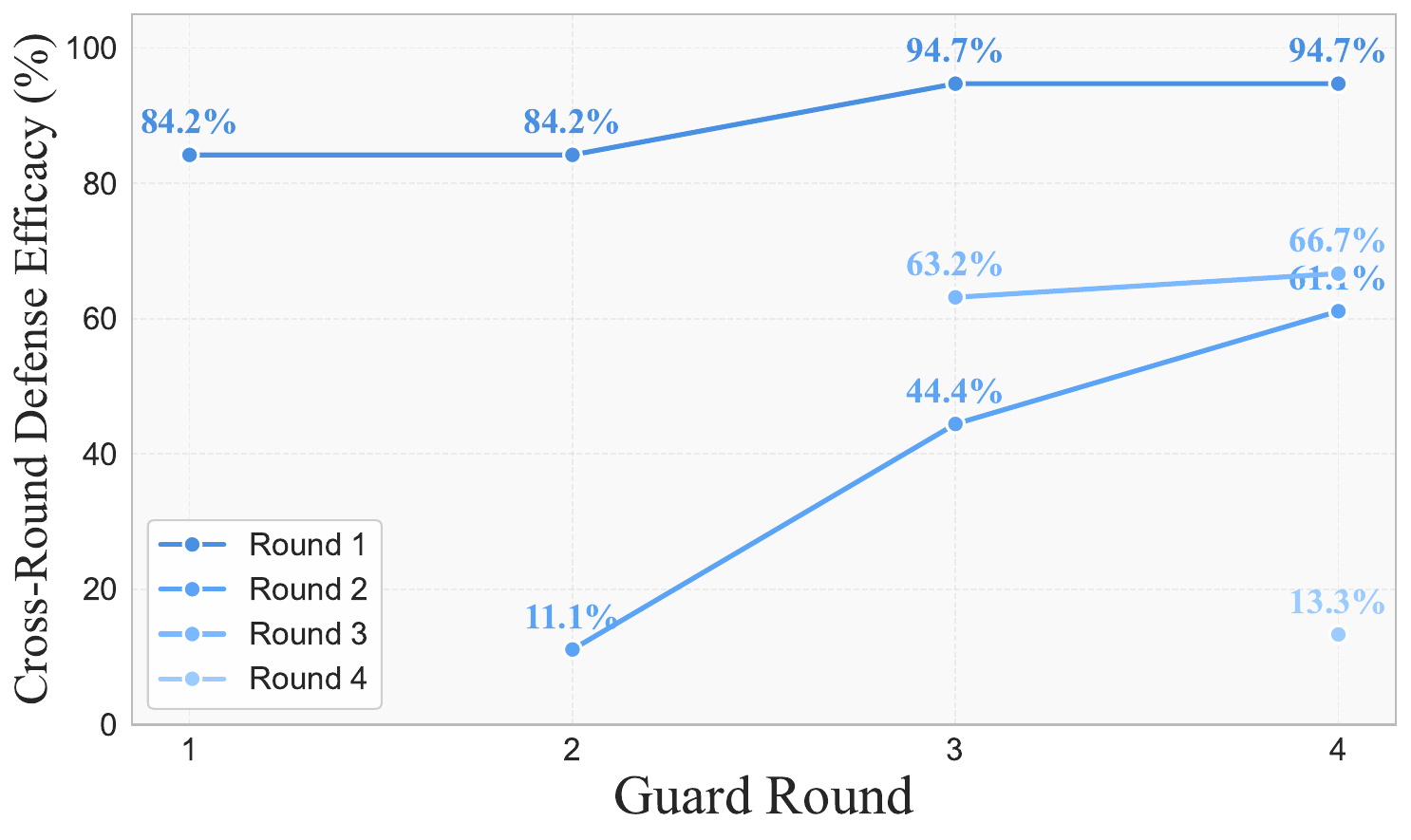}
    \caption{Cross-Round Defense Efficacy (CRDE) for exp. 2. Each line shows that newer guardrails (e.g., Round 4) perform better against attacks from all rounds, including older ones (e.g., Round 1), indicating no catastrophic forgetting.}
    \label{fig:crde}
\end{figure}

\begin{figure}[t!]
    \centering
    \includegraphics[width=1.0\linewidth]{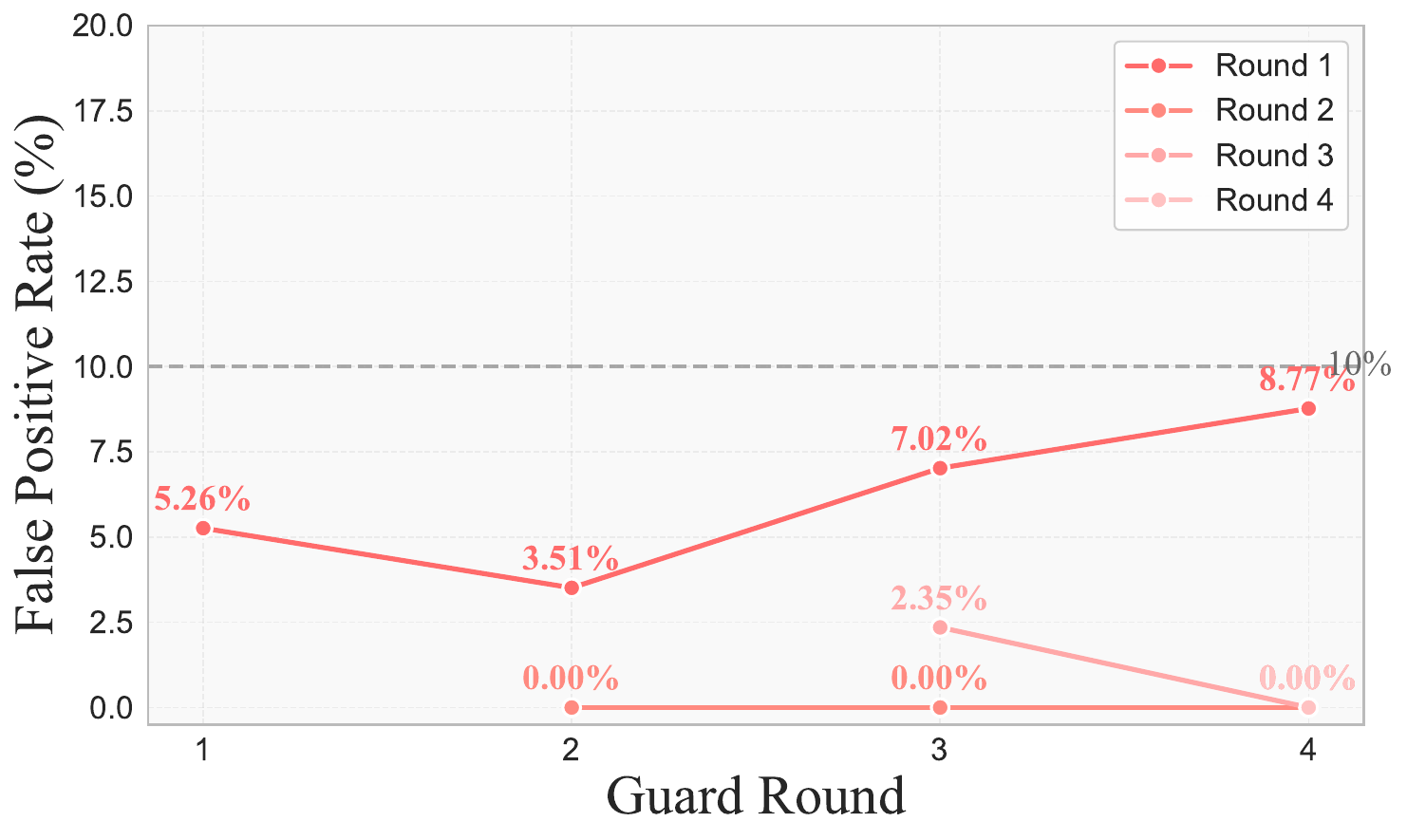}
    \caption{False Positive Rate (FPR) for exp. 2. The FPR remains consistently low across all rounds, indicating minimal over-interception of benign prompts.}
    \label{fig:fpr}
\end{figure}

\textbf{2) The Blue Team learns generalizable rules without catastrophic forgetting.}
A crucial aspect of iterative learning is avoiding overfitting to the most recent attacks. Figure \ref{fig:crde} illustrates the Cross-Round Defense Efficacy (CRDE), showing how guardrails from later rounds perform against attacks from earlier rounds. Each line represents the performance of a progressively updated guardrail (from Guard Round 1 to 4) tested against a fixed set of historical attacks (from Round 1, 2, 3, or 4). A clear upward trend is observable across all lines, demonstrating two key findings. Firstly, the guardrail's ability to defend against new attacks improves with each iteration. Secondly, and more importantly, its defense performance against older attack sets (e.g., the Guard Round 4's performance on Round 1 attacks) does not degrade but consistently improves. This indicates our iterative augmentation mechanism successfully avoids catastrophic forgetting. Furthermore, as shown in Figure \ref{fig:fpr}, the False Positive Rate (FPR) remains at a very low level, demonstrating that the enhanced security does not come at the cost of excessive over-interception.

\begin{figure}[t!]
    \centering
    \includegraphics[width=1.0\linewidth]{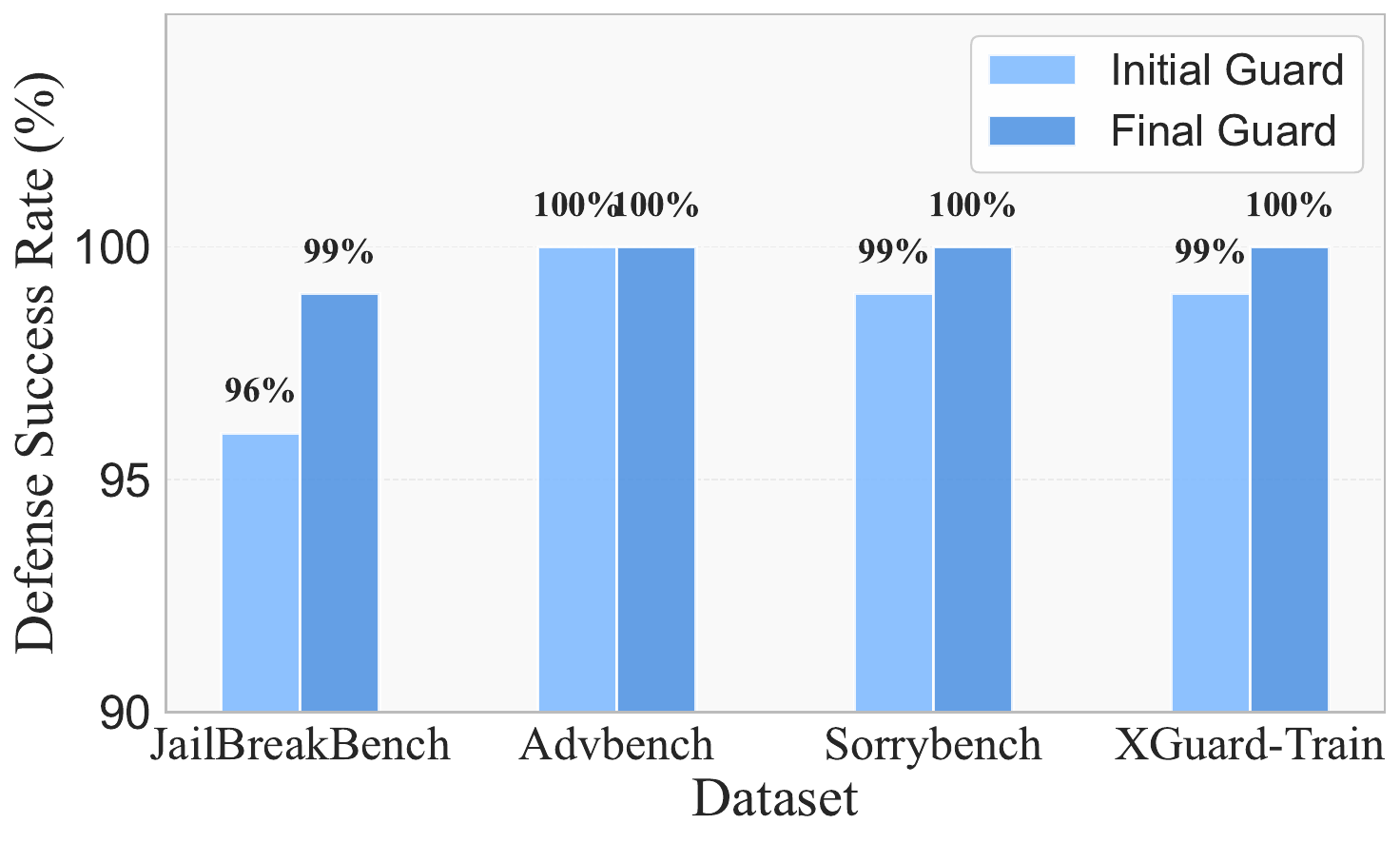}
    \caption{Out-of-Domain performance comparison between the initial and final guardrail. The final guardrail shows significantly improved defensive performance on four unseen external benchmarks.}
    \label{fig:ood}
\end{figure}

\textbf{3) The evolved guardrail demonstrates strong out-of-domain generalization.} To verify that the learned defensive capabilities generalize beyond the specific attack patterns seen during interaction, we evaluated the initial and final guardrails on 100 randomly selected prompts from four external, unseen jailbreak benchmarks. As shown in Figure \ref{fig:ood}, the Final Guard consistently outperforms the Initial Guard across all external datasets (JailBreakBench \citep{chao2024jailbreakbenchopenrobustnessbenchmark}, AdvBench \citep{zou2023universal}, SorryBench \citep{xie2024sorry}, and XGuard-Train \citep{upadhayay2025xguardmultilingualguardagent}), with its defense success rate improving from 96\% to 99\% on JailBreakBench and achieving a perfect 100\% on AdvBench. This indicates that the adversarial iterations compelled the Blue Team to learn more fundamental and robust principles of defense, enhancing its resilience against a broader spectrum of unseen jailbreak attacks.

\section{Discussion}
\label{sec:discussion}

Our findings substantiate the efficacy of the \textbf{RvB} framework, offering improvements that stem directly from the game-theoretic structure rather than mere model scaling. We analyze the mechanisms driving these results below.

\paragraph{Convergence to Practical Equilibrium.}
The adversarial interaction drives the system toward a \textit{practical equilibrium} where defensive coverage matches the high-probability attack manifold. As the Red Team's belief entropy ($H(b_k)$) decreases, its strategy shifts from broad exploration to focused exploitation. The resulting convergence of high Defense Success Rates with escalating Attack Costs suggests that further exploitation becomes computationally prohibitive. While a formal Nash Equilibrium in the high-dimensional discrete space of natural language remains theoretically elusive, the monotonic state progression confirms that offensive and defensive utilities balance at a robust security standard.

\paragraph{Interaction-Driven Generalization.}
The strong performance on out-of-domain attacks (Figure~\ref{fig:ood}) indicates that robustness emerges from the adversarial dynamic rather than latent model knowledge alone. In this zero-sum loop, the Red Team actively probes for blind spots, functioning as a continuous regularizer. This pressure penalizes superficial pattern matching, forcing the Blue Team to induce abstract, invariant security principles—such as detecting malicious intent structures over specific keywords. The interaction effectively synthesizes a curriculum of ``hard negatives'' from the tail of the vulnerability distribution, driving generalization without explicit supervision.

\paragraph{Operational Viability via Semantic Verification.}
Crucially, the adversarial pairing resolves the ``destructive remediation'' bottleneck often found in autonomous software engineering. The near-zero Service Disruption Rate (SDR) confirms that the Red Team serves as a stringent semantic verifier. By coupling remediation with immediate, targeted exploitation, the framework imposes a check-and-balance dynamic absent in cooperative systems, ensuring that security hardening does not compromise system availability.

\section{Conclusion}
\label{sec:conclusion}
In this work, we present a unified \textbf{Red Team vs. Blue Team (RvB)} framework that automates the continuous hardening of AI systems. By formalizing security operations as a sequential, imperfect-information game, we bridge the gap between offensive discovery and defensive mitigation, enabling a self-reinforcing loop of capability enhancement without model tuning.

Experiments on cyber and content security demonstrate that this paradigm shifts security automation from static, post-hoc patching to iterative adversarial adaptation. The framework not only successfully repairs vulnerabilities with high precision and minimal service disruption, but also produces defensive guardrails with strong out-of-distribution generalization. These findings suggest that the most effective way to secure foundation models is to subject them to the adversarial pressures they face in real-world scenarios, governed by a structured game-theoretic architecture.

Future work will explore the theoretical bounds of this convergence in multi-modal environments and investigate the application of this framework to more complex, long-horizon security tasks.

\section*{Limitations}
The efficacy of the RvB framework relies intrinsically on the reasoning capabilities of the underlying foundation models. Regarding evaluation scope, our current experiments utilized a representative but constrained set of web vulnerabilities. Rigorous testing against larger-scale, diverse benchmarks is required to further verify generalizability. We also note that the turn-based simulation serves as a controlled abstraction, effectively simplifying the asynchronous dynamics inherent in real-world security operations.

\section*{Ethics Statement}
This work involves the development of automated agents capable of executing software exploits and generating adversarial jailbreaks. We acknowledge the dual-use nature of such technologies; while intended for system hardening, the offensive capabilities of the Red Team agent could theoretically be repurposed for malicious attacks. To mitigate these risks, we strictly adhered to the following safety protocols:

\paragraph{Controlled Environments.} All cybersecurity experiments were conducted within isolated, network-gated Docker containers with no internet access beyond the local simulation loop. This ensures that the automated exploit generation cannot propagate to external systems or critical infrastructure.

\paragraph{Defensive Focus.} The primary objective of the RvB framework is to minimize the epistemic uncertainty of the defender (Blue Team). The offensive agent functions solely as a diagnostic tool to drive this defensive adaptation. We do not release offensive agent that function independently of the hardening loop.

\paragraph{Responsible Disclosure and Data.} For the content security tasks, we utilized public benchmarks (e.g., HarmBench) and strictly monitored the generated outputs. No new private datasets of harmful content were distributed. Our code release includes safeguards to prevent the standalone deployment of the Red Team agent for malicious probing. We believe that documenting these automated red-teaming dynamics is essential for the community to develop robust defenses against emerging autonomous threats.

\bibliography{main}

@misc{mayoralvilches2025caiopenbugbountyready,
      title={CAI: An Open, Bug Bounty-Ready Cybersecurity AI}, 
      author={Víctor Mayoral-Vilches and Luis Javier Navarrete-Lozano and María Sanz-Gómez and Lidia Salas Espejo and Martiño Crespo-Álvarez and Francisco Oca-Gonzalez and Francesco Balassone and Alfonso Glera-Picón and Unai Ayucar-Carbajo and Jon Ander Ruiz-Alcalde and Stefan Rass and Martin Pinzger and Endika Gil-Uriarte},
      year={2025},
      eprint={2504.06017},
      archivePrefix={arXiv},
      primaryClass={cs.CR},
      url={https://arxiv.org/abs/2504.06017}, 
}

@misc{shen2025pentestagentincorporatingllmagents,
      title={PentestAgent: Incorporating LLM Agents to Automated Penetration Testing}, 
      author={Xiangmin Shen and Lingzhi Wang and Zhenyuan Li and Yan Chen and Wencheng Zhao and Dawei Sun and Jiashui Wang and Wei Ruan},
      year={2025},
      eprint={2411.05185},
      archivePrefix={arXiv},
      primaryClass={cs.CR},
      url={https://arxiv.org/abs/2411.05185}, 
}

@inproceedings{Castro_2025,
   title={Large Language Models are Autonomous Cyber Defenders},
   url={http://dx.doi.org/10.1109/CAI64502.2025.00195},
   DOI={10.1109/cai64502.2025.00195},
   booktitle={2025 IEEE Conference on Artificial Intelligence (CAI)},
   publisher={IEEE},
   author={Castro, Sebastián R. and Campbell, Roberto and Lau, Nancy and Villalobos, Octavio and Duan, Jiaqi and Cardenas, Alvaro A.},
   year={2025},
   month=may, pages={1125–1132} }

@inproceedings{yang2024sweagent,
  title={{SWE}-agent: Agent-Computer Interfaces Enable Automated Software Engineering},
  author={John Yang and Carlos E Jimenez and Alexander Wettig and Kilian Lieret and Shunyu Yao and Karthik R Narasimhan and Ofir Press},
  booktitle={The Thirty-eighth Annual Conference on Neural Information Processing Systems},
  year={2024},
  url={https://arxiv.org/abs/2405.15793}
}

@article{dorri2018multi,
  title={Multi-agent systems: A survey},
  author={Dorri, Ali and Kanhere, Salil S and Jurdak, Raja},
  journal={Ieee Access},
  volume={6},
  pages={28573--28593},
  year={2018},
  publisher={IEEE}
}

@article{tao2024magis,
  title={Magis: Llm-based multi-agent framework for github issue resolution},
  author={Tao, Wei and Zhou, Yucheng and Wang, Yanlin and Zhang, Wenqiang and Zhang, Hongyu and Cheng, Yu},
  journal={Advances in Neural Information Processing Systems},
  volume={37},
  pages={51963--51993},
  year={2024}
}

@article{wang2024multi,
  title={Multi-agent collaboration framework for recommender systems},
  author={Wang, Zhefan and Yu, Yuanqing and Zheng, Wendi and Ma, Weizhi and Zhang, Min},
  journal={arXiv preprint arXiv:2402.15235
        
        
        
        
        
        
        
        
        
        
        
        
        
        
        
        
        
        
        
        
        
        },
  year={2024}
}

@article{ishibashi2024self,
  title={Self-organized agents: A llm multi-agent framework toward ultra large-scale code generation and optimization},
  author={Ishibashi, Yoichi and Nishimura, Yoshimasa},
  journal={arXiv preprint arXiv:2404.02183
        
        
        
        
        
        
        
        
        
        
        
        
        
        
        
        
        
        
        
        
        
        
        
        
        
        
        
        
        
        
        
        
        
        },
  year={2024}
}

@article{yang2025agentnet,
  title={Agentnet: Decentralized evolutionary coordination for llm-based multi-agent systems},
  author={Yang, Yingxuan and Chai, Huacan and Shao, Shuai and Song, Yuanyi and Qi, Siyuan and Rui, Renting and Zhang, Weinan},
  journal={arXiv preprint arXiv:2504.00587
        
        
        
        
        
        
        
        
        
        
        
        
        
        
        
        
        
        },
  year={2025}
}

@article{jha2025cross,
  title={Cross-environment Cooperation Enables Zero-shot Multi-agent Coordination},
  author={Jha, Kunal and Carvalho, Wilka and Liang, Yancheng and Du, Simon S and Kleiman-Weiner, Max and Jaques, Natasha},
  journal={arXiv preprint arXiv:2504.12714
        
        
        
        
        
        
        
        
        
        
        
        
        
        
        
        
        
        
        
        
        
        
        
        
        
        
        
        
        
        
        
  
        
        },
  year={2025}
}

@article{lee2024prompt,
  title={Prompt infection: Llm-to-llm prompt injection within multi-agent systems},
  author={Lee, Donghyun and Tiwari, Mo},
  journal={arXiv preprint arXiv:2410.07283
        
        
        
        
        
        
        
        
        
        
        
        
        
        
        
        
        
        
        
        
        
        
        
        
        
        
        
        
        
        
        
        
        
        
        
        
        
        
        
        

        },
  year={2024}
}

@inproceedings{he2025red,
  title={Red-teaming llm multi-agent systems via communication attacks},
  author={He, Pengfei and Lin, Yuping and Dong, Shen and Xu, Han and Xing, Yue and Liu, Hui},
  booktitle={Findings of the Association for Computational Linguistics: ACL 2025},
  pages={6726--6747},
  year={2025}
}

@article{arora2025exposing,
  title={Exposing Weak Links in Multi-Agent Systems under Adversarial Prompting},
  author={Arora, Nirmit and Joel, Sathvik and Kavathekar, Ishan and Gandhi, Rohan and Pandya, Yash and Ganu, Tanuja and Kanade, Aditya and Nambi, Akshay and others},
  journal={arXiv preprint arXiv:2511.10949},
  year={2025}
}

@article{fan2025peerguard,
  title={PeerGuard: Defending Multi-Agent Systems Against Backdoor Attacks Through Mutual Reasoning},
  author={Fan, Falong and Li, Xi},
  journal={arXiv preprint arXiv:2505.11642
        
        
        
        
        
        
        
        
        
        
        
        
        
        
        
        
        
        
        
        
        
        
        
        
        
        
        
        
        
        },
  year={2025}
}

@article{li2025security,
  title={Security concerns for large language models: A survey},
  author={Li, Miles Q and Fung, Benjamin CM},
  journal={Journal of Information Security and Applications},
  volume={95},
  pages={104284},
  year={2025},
  publisher={Elsevier}
}

@article{mazeika2024harmbench,
  title={Harmbench: A standardized evaluation framework for automated red teaming and robust refusal},
  author={Mazeika, Mantas and Phan, Long and Yin, Xuwang and Zou, Andy and Wang, Zifan and Mu, Norman and Sakhaee, Elham and Li, Nathaniel and Basart, Steven and Li, Bo and others},
  journal={arXiv preprint arXiv:2402.04249
        
        
        
        
        
        
        
        
        
        
        
        
        
        
        
        
        
        
        
        
        
        
        
        
        
        
        
        
        
        
        
        
        
        
        
        
        
        
        
        },
  year={2024}
}

@article{zeng2024autodefense,
  title={Autodefense: Multi-agent llm defense against jailbreak attacks},
  author={Zeng, Yifan and Wu, Yiran and Zhang, Xiao and Wang, Huazheng and Wu, Qingyun},
  journal={arXiv preprint arXiv:2403.04783
        
        
        
        
        
        
        
        
        
        
        
        
        
        
        
        
        
        
        
        
        
        
        
        
        
        
        
        
        
        
        
        
        
        
        
        
        
        
        
        
        
        },
  year={2024}
}

@article{zou2025security,
  title={Security challenges in ai agent deployment: Insights from a large scale public competition},
  author={Zou, Andy and Lin, Maxwell and Jones, Eliot and Nowak, Micha and Dziemian, Mateusz and Winter, Nick and Grattan, Alexander and Nathanael, Valent and Croft, Ayla and Davies, Xander and others},
  journal={arXiv preprint arXiv:2507.20526
        
        
        
        
        
        
        
        
        
        
        
        
        
        
        
        
        
        
        
        
        
        
        
        
        
        
        
        
        
        
        
        
        
        
        
        
        
        
        
        
        
        
        
        },
  year={2025}
}

@misc{chao2024jailbreakbenchopenrobustnessbenchmark,
      title={JailbreakBench: An Open Robustness Benchmark for Jailbreaking Large Language Models}, 
      author={Patrick Chao and Edoardo Debenedetti and Alexander Robey and Maksym Andriushchenko and Francesco Croce and Vikash Sehwag and Edgar Dobriban and Nicolas Flammarion and George J. Pappas and Florian Tramer and Hamed Hassani and Eric Wong},
      year={2024},
      eprint={2404.01318},
      archivePrefix={arXiv},
      primaryClass={cs.CR},
      url={https://arxiv.org/abs/2404.01318}, 
}

@article{xiong2025cop,
  title={CoP: Agentic Red-teaming for Large Language Models using Composition of Principles},
  author={Xiong, Chen and Chen, Pin-Yu and Ho, Tsung-Yi},
  journal={arXiv preprint arXiv:2506.00781
        
        
        
        
        
        
        
        
        
        
        
        
        
        
        
        
        
        
        
        
        
        
        
        
        
        
        
        
        
        
        
        
        
        
        
        
        
        
        
        
        
        },
  year={2025}
}

@article{huynh2025understanding,
  title={Understanding LLM Agent Behaviours via Game Theory: Strategy Recognition, Biases and Multi-Agent Dynamics},
  author={Huynh, Trung-Kiet and Dao-Sy, Duy-Minh and Cao, Thanh-Bang and Le, Phong-Hao and Nguyen, Hong-Dan and Nguyen-Lam, Phu-Quy and Nguyen-Vo, Minh-Luan and Pham, Hong-Phat and Pham, Phu-Hoa and Than, Thien-Kim and others},
  journal={arXiv preprint arXiv:2512.07462
        
        
        
        
        
        
        
        
        
        
        
        
        
        
        
        
        
        
        
        
        
        
        
        
        
        
        
        
        
        
        
        
        
        },
  year={2025}
}

@inproceedings{rebedea2023nemo,
  title={Nemo guardrails: A toolkit for controllable and safe llm applications with programmable rails},
  author={Rebedea, Traian and Dinu, Razvan and Sreedhar, Makesh Narsimhan and Parisien, Christopher and Cohen, Jonathan},
  booktitle={Proceedings of the 2023 conference on empirical methods in natural language processing: system demonstrations},
  pages={431--445},
  year={2023}
}

@article{zou2023universal,
  title={Universal and transferable adversarial attacks on aligned language models},
  author={Zou, Andy and Wang, Zifan and Carlini, Nicholas and Nasr, Milad and Kolter, J Zico and Fredrikson, Matt},
  journal={arXiv preprint arXiv:2307.15043
        
        
        
        
        
        
        
        
        
        
        
        
        
        
        
        },
  year={2023}
}

@article{xie2024sorry,
  title={Sorry-bench: Systematically evaluating large language model safety refusal},
  author={Xie, Tinghao and Qi, Xiangyu and Zeng, Yi and Huang, Yangsibo and Sehwag, Udari Madhushani and Huang, Kaixuan and He, Luxi and Wei, Boyi and Li, Dacheng and Sheng, Ying and others},
  journal={arXiv preprint arXiv:2406.14598
        
        
        
        
        
        
        
        
        
        
        
        
        
        
        
        
        
        
        
        
        
        
        
        
        
        
        
        
        
        
        
        
        
        
        
        
        
        
        
        
        
        
        
        
        
        },
  year={2024}
}

@misc{upadhayay2025xguardmultilingualguardagent,
      title={X-Guard: Multilingual Guard Agent for Content Moderation}, 
      author={Bibek Upadhayay and Vahid Behzadan and Ph. D},
      year={2025},
      eprint={2504.08848},
      archivePrefix={arXiv},
      primaryClass={cs.CR},
      url={https://arxiv.org/abs/2504.08848}, 
}

@article{ying2025reasoning,
  title={Reasoning-augmented conversation for multi-turn jailbreak attacks on large language models},
  author={Ying, Zonghao and Zhang, Deyue and Jing, Zonglei and Xiao, Yisong and Zou, Quanchen and Liu, Aishan and Liang, Siyuan and Zhang, Xiangzheng and Liu, Xianglong and Tao, Dacheng},
  journal={arXiv preprint arXiv:2502.11054
        
        
        
        
        
        
        
        
        
        
        
        
        
        
        
        
        
        
        
        
        
        
        
        
        
        
        
        
        
        
        
        
        
        
        
        
        
        
        
        },
  year={2025}
}

@article{liu2025automated,
  title={An Automated Framework for Strategy Discovery, Retrieval, and Evolution in LLM Jailbreak Attacks},
  author={Liu, Xu and Chen, Yan and Ling, Kan and Zhu, Yichi and Zhang, Hengrun and Fan, Guisheng and Yu, Huiqun},
  journal={arXiv preprint arXiv:2511.02356
        
        
        
        
        
        
        
        
        
        
        
        
        
        
        
        
        
        
        
        
        
        
        
        
        
        
        
        
        
        
        
        
        
        
        
        
        
        
        
        
        
        
        
        },
  year={2025}
}

@article{rahman2025x,
  title={X-teaming: Multi-turn jailbreaks and defenses with adaptive multi-agents},
  author={Rahman, Salman and Jiang, Liwei and Shiffer, James and Liu, Genglin and Issaka, Sheriff and Parvez, Md Rizwan and Palangi, Hamid and Chang, Kai-Wei and Choi, Yejin and Gabriel, Saadia},
  journal={arXiv preprint arXiv:2504.13203
        
        
        
        
        
        
        
        
        
        
        
        
        
        
        
        
        
        
        
        
        
        
        
        
        
        
        
        
        
        
        
        
        
        
        
        },
  year={2025}
}

@misc{zhou2024robustpromptoptimizationdefending,
      title={Robust Prompt Optimization for Defending Language Models Against Jailbreaking Attacks}, 
      author={Andy Zhou and Bo Li and Haohan Wang},
      year={2024},
      eprint={2401.17263},
      archivePrefix={arXiv},
      primaryClass={cs.LG},
      url={https://arxiv.org/abs/2401.17263}, 
}

@misc{xiong2025defensivepromptpatchrobust,
      title={Defensive Prompt Patch: A Robust and Interpretable Defense of LLMs against Jailbreak Attacks}, 
      author={Chen Xiong and Xiangyu Qi and Pin-Yu Chen and Tsung-Yi Ho},
      year={2025},
      eprint={2405.20099},
      archivePrefix={arXiv},
      primaryClass={cs.CR},
      url={https://arxiv.org/abs/2405.20099}, 
}

@misc{hoover2025dynaguarddynamicguardianmodel,
      title={DynaGuard: A Dynamic Guardian Model With User-Defined Policies}, 
      author={Monte Hoover and Vatsal Baherwani and Neel Jain and Khalid Saifullah and Joseph Vincent and Chirag Jain and Melissa Kazemi Rad and C. Bayan Bruss and Ashwinee Panda and Tom Goldstein},
      year={2025},
      eprint={2509.02563},
      archivePrefix={arXiv},
      primaryClass={cs.LG},
      url={https://arxiv.org/abs/2509.02563}, 
}

@misc{yang2025retrievalaugmenteddefenseadaptivecontrollable,
      title={Retrieval-Augmented Defense: Adaptive and Controllable Jailbreak Prevention for Large Language Models}, 
      author={Guangyu Yang and Jinghong Chen and Jingbiao Mei and Weizhe Lin and Bill Byrne},
      year={2025},
      eprint={2508.16406},
      archivePrefix={arXiv},
      primaryClass={cs.CR},
      url={https://arxiv.org/abs/2508.16406}, 
}

@misc{zhu2025reasoningtodefendsafetyawarereasoningdefend,
      title={Reasoning-to-Defend: Safety-Aware Reasoning Can Defend Large Language Models from Jailbreaking}, 
      author={Junda Zhu and Lingyong Yan and Shuaiqiang Wang and Dawei Yin and Lei Sha},
      year={2025},
      eprint={2502.12970},
      archivePrefix={arXiv},
      primaryClass={cs.CL},
      url={https://arxiv.org/abs/2502.12970}, 
}

@misc{yang2025enhancingmodeldefensejailbreaks,
      title={Enhancing Model Defense Against Jailbreaks with Proactive Safety Reasoning}, 
      author={Xianglin Yang and Gelei Deng and Jieming Shi and Tianwei Zhang and Jin Song Dong},
      year={2025},
      eprint={2501.19180},
      archivePrefix={arXiv},
      primaryClass={cs.CR},
      url={https://arxiv.org/abs/2501.19180}, 
}

@misc{deng2024pentestgptllmempoweredautomaticpenetration,
      title={PentestGPT: An LLM-empowered Automatic Penetration Testing Tool}, 
      author={Gelei Deng and Yi Liu and Víctor Mayoral-Vilches and Peng Liu and Yuekang Li and Yuan Xu and Tianwei Zhang and Yang Liu and Martin Pinzger and Stefan Rass},
      year={2024},
      eprint={2308.06782},
      archivePrefix={arXiv},
      primaryClass={cs.SE},
      url={https://arxiv.org/abs/2308.06782}, 
}

@misc{henke2025autopentestenhancingvulnerabilitymanagement,
      title={AutoPentest: Enhancing Vulnerability Management With Autonomous LLM Agents}, 
      author={Julius Henke},
      year={2025},
      eprint={2505.10321},
      archivePrefix={arXiv},
      primaryClass={cs.CR},
      url={https://arxiv.org/abs/2505.10321}, 
}

@misc{deng2023largelanguagemodelszeroshot,
      title={Large Language Models are Zero-Shot Fuzzers: Fuzzing Deep-Learning Libraries via Large Language Models}, 
      author={Yinlin Deng and Chunqiu Steven Xia and Haoran Peng and Chenyuan Yang and Lingming Zhang},
      year={2023},
      eprint={2212.14834},
      archivePrefix={arXiv},
      primaryClass={cs.SE},
      url={https://arxiv.org/abs/2212.14834}, 
}

@inproceedings{Xia_2024, series={ICSE ’24},
   title={Fuzz4All: Universal Fuzzing with Large Language Models},
   url={http://dx.doi.org/10.1145/3597503.3639121},
   DOI={10.1145/3597503.3639121},
   booktitle={Proceedings of the IEEE/ACM 46th International Conference on Software Engineering},
   publisher={ACM},
   author={Xia, Chunqiu Steven and Paltenghi, Matteo and Le Tian, Jia and Pradel, Michael and Zhang, Lingming},
   year={2024},
   month=apr, pages={1–13},
   collection={ICSE ’24} }

@misc{jin2023inferfixendtoendprogramrepair,
      title={InferFix: End-to-End Program Repair with LLMs}, 
      author={Matthew Jin and Syed Shahriar and Michele Tufano and Xin Shi and Shuai Lu and Neel Sundaresan and Alexey Svyatkovskiy},
      year={2023},
      eprint={2303.07263},
      archivePrefix={arXiv},
      primaryClass={cs.SE},
      url={https://arxiv.org/abs/2303.07263}, 
}

@article{de_Fitero_Dominguez_2024,
   title={Enhanced automated code vulnerability repair using large language models},
   volume={138},
   ISSN={0952-1976},
   url={http://dx.doi.org/10.1016/j.engappai.2024.109291},
   DOI={10.1016/j.engappai.2024.109291},
   journal={Engineering Applications of Artificial Intelligence},
   publisher={Elsevier BV},
   author={de-Fitero-Dominguez, David and Garcia-Lopez, Eva and Garcia-Cabot, Antonio and Martinez-Herraiz, Jose-Javier},
   year={2024},
   month=dec, pages={109291} }

@misc{yang2024sweagentagentcomputerinterfacesenable,
      title={SWE-agent: Agent-Computer Interfaces Enable Automated Software Engineering}, 
      author={John Yang and Carlos E. Jimenez and Alexander Wettig and Kilian Lieret and Shunyu Yao and Karthik Narasimhan and Ofir Press},
      year={2024},
      eprint={2405.15793},
      archivePrefix={arXiv},
      primaryClass={cs.SE},
      url={https://arxiv.org/abs/2405.15793}, 
}

\appendix
\section{Experiment Details of The Attacker vs. Defender in Cyber Security}
\label{sec:appendix_exp1}
\subsection{Red Team Setting}
The architecture of the red team agent (CAI) consists of three core components: a \textit{Planner} responsible for deducing attack paths, an \textit{Executor} for executing specific Bash/MCP commands, and a \textit{Reporter} for summarizing attack outcomes. During the offensive process, the red team first performs passive reconnaissance on the target environment. The \textit{Planner} generates attack hypotheses based on system feedback, after which the \textit{Executor} invokes tools to conduct active probing and payload delivery. Upon a successful exploit, the agent maintains access and triggers the \textit{Reporter} to generate a vulnerability report containing reproduction steps. For this experiment, we selected \texttt{gemini-3.0-flash} as the backbone model. The maximum number of interaction turns was set to 30 to ensure sufficient probing depth.

\subsection{Blue Team Setting}
The blue team's Mini-SWE-Agent is designed to simulate the remediation workflow of a security engineer. Its inputs are the vulnerability report generated by the red team and the project's source code. The blue team's workflow comprises three stages: \textit{fault localization}, \textit{patch generation}, and \textit{regression verification}. First, the agent analyzes the codebase based on the vulnerability report to locate the vulnerable PHP files. Next, it generates a patch in \texttt{git diff} format and applies it to the environment. Finally, it restarts the Docker container to verify service availability and confirm the mitigation of the vulnerability. To ensure the accuracy of code modifications and prevent disruption of existing business logic, the blue team model requires a high level of code comprehension and generation capability; thus, we also selected \texttt{gemini-3.0-flash} as its underlying model. If the initial repair fails, the blue team is permitted a maximum of 3 retry attempts to correct the patch.

\subsection{Metric Definitions for Cyber Security}
\label{sec:metric_definitions}

In the context of the Cyber Security (Code Hardening) task, let $\mathcal{D} = \{x_1, x_2, \dots, x_N\}$ denote the set of vulnerability test cases. For each test case $x_i$ at a given interaction round, we define the outcome state as a tuple $(r_{att}^{(i)}, r_{reg}^{(i)})$, where:
\begin{itemize}
    \item $r_{att}^{(i)} \in \{0, 1\}$ denotes the outcome of the Red Team's attack. $r_{att}^{(i)}=1$ indicates a successful breach (exploit succeeded), while $0$ indicates a failed attack (defense succeeded).
    \item $r_{reg}^{(i)} \in \{0, 1\}$ denotes the outcome of the service regression test. $r_{reg}^{(i)}=1$ indicates the service remains functional, while $0$ indicates a service disruption (e.g., HTTP 500 error or missing interface).
\end{itemize}

\paragraph{Defensive Metrics: DSR, TDSR, and FDSR}

The \textbf{Defense Success Rate (DSR)} represents the overall ratio of cases where the Blue Team successfully fixes the system.

\noindent \textbf{True Defense Success Rate (TDSR)}: The proportion of cases where the vulnerability is mitigated \textit{and} the service integrity is preserved.
\begin{equation}
    \text{DSR} =\text{TDSR} = \frac{1}{N} \sum_{i=1}^{N} \mathbb{I}(r_{att}^{(i)} = 0 \land r_{reg}^{(i)} = 1)
\end{equation}

\noindent \textbf{Fake Defense Success Rate (FDSR)}: The proportion of cases where the attack fails solely because the defensive patch rendered the service non-functional.
\begin{equation}
    \text{FDSR} = \frac{1}{N} \sum_{i=1}^{N} \mathbb{I}(r_{att}^{(i)} = 0)
\end{equation}

\noindent \textbf{Service Disruption Rate (SDR)}: A measure of availability loss caused by defensive over-optimization. It denotes the frequency with which valid user requests are blocked or service interfaces become unresponsive as a direct result of the applied security patches.
\begin{equation}
    \text{SDR} = \frac{1}{N} \sum_{i=1}^{N} \mathbb{I}(r_{att}^{(i)} = 0 \land r_{reg}^{(i)} = 0)
\end{equation}

From the equations above, it holds that $\text{SDR} = \text{FDSR}-\text{TDSR} $. A robust automated defense agent should maximize TDSR while maintaining SDR near zero.

\paragraph{Offensive Metric: Attack Success Count (ASC)}

We utilize \textbf{Attack Success Count (ASC)} to measure the diversity and complexity of the attack vectors discovered by the Red Team. Let $\mathcal{E}_k$ be the set of unique, valid exploit payloads successfully executed by the Red Team up to round $k$. The ASC is defined as the cardinality of this cumulative set:
\begin{equation}
\begin{split}
    \text{ASC}_k &= |\mathcal{E}_k| \\
    &= \left| \bigcup_{t=1}^{k} \{e \mid e \in \text{Payloads}_t, \text{IsSuccess}(e)\} \right|
\end{split}
\end{equation}
Higher ASC values indicate that the Red Team is capable of evolving its strategy to uncover deeper logic vulnerabilities as the environment hardens.

\subsection{Detailed Procedure}

\begin{enumerate}
    \item \textbf{Attack Generation (Red Team):} 
    The Red agent initiates the cycle by interacting with the current environment ($t$ Env). It performs reconnaissance and exploitation to identify security flaws. Upon successful compromise, it produces a structured \textit{Attack Log} containing the vulnerable file path, the snippet of problematic code, and the root cause of the bug. A representative log entry is formatted as follows:
    
\begin{center}
\fbox{\parbox{0.95\linewidth}{%
    \small\ttfamily %
    \{ \\
    \textbf{"file"}: "php\_action/removeOrder.php", \\
    \textbf{"code"}: "\$orderId= \$\_GET['id'];\textbackslash n if(\$orderId) \{ \textbackslash n \$sql = \textbackslash"UPDATE orders SET order\_status = 2 WHERE order\_id = \{\$orderId\}\textbackslash"; ... \}", \\
    \textbf{"bug"}: "SQL Injection via 'id' parameter in GET request. The variable \$orderId is interpolated without sanitization.", \\
    \textbf{"payload"}: "id=1 OR 1=1" \\
    \}%
}}
\end{center}
    
    This structured output details the exploit methodology and reproduction steps, serving as the ground truth for the subsequent fix.

    \item \textbf{Remediation (Blue Team):} 
    Acting on the information provided in the \textit{Attack Log}, the Blue agent analyzes the codebase to localize the fault. It then synthesizes a code patch and performs a \textit{Fix Act} to repair the vulnerability.

    \item \textbf{State Update \& Verification:} 
    The generated patch is applied to the Docker container, transitioning the system to \textit{Target State $t+1$}. Finally, a \textit{Vuln Check} is executed to verify the effectiveness of the fix and ensure no regression errors were introduced. If the vulnerability persists, the Blue agent is triggered to refine its patch in subsequent iterations.
\end{enumerate}

\section{Experiment Details of The Jailbreaker vs. Guardrail in Content Security}
\label{sec:appendix_exp2_details}

\subsection{Agent Architecture and Model Selection}
The Red and Blue teams were designed with specialized components and models to fulfill their respective roles in the adversarial framework.

\paragraph{Red Team Components} The Red Team agent, based on the Composition-of-Principles (CoP) framework, includes four main LLM-driven components:
\begin{itemize}
    \item \textbf{Target Model (\textit{target\_model}):} The model to be attacked, protected by the Blue Team's guardrail. To better observe the iterative improvement of the guardrail, this model was selected to have relatively weak intrinsic safety features but adequate question-answering capabilities. We chose \textbf{Qwen2-7B-Instruct}.
    \item \textbf{Helper Model (\textit{helper\_model}):} Responsible for generating diverse attack strategies. This role requires high capability without being prone to refusal. We chose \textbf{DeepSeek-V3}.
    \item \textbf{Attack Model (\textit{attack\_model}):} Guided by the strategies from the helper model, this component generates the actual attack prompts.
    \item \textbf{Judge Model (\textit{judge\_model}):} Evaluates the success of an attack by checking if the target model's response aligns with the harmful intent, while also monitoring for topic drift.
\end{itemize}
For the \textit{attack\_model} and \textit{judge\_model}, we required highly capable open-source models that could reliably follow complex instructions. We chose \textbf{Qwen3-32B} for both roles.

\paragraph{Blue Team Components} The Blue Team agent, based on NeMo Guardrails, consists of the following:
\begin{itemize}
    \item \textbf{Target Model (\textit{target\_model}):} The same \textbf{Qwen2-7B-Instruct} used by the Red Team.
    \item \textbf{Guard Model (\textit{guard\_model}):} Used during the defense phase to classify incoming prompts based on the current rule set.
    \item \textbf{Patch Model (\textit{patch\_model}):} Used to analyze successful attack prompts and augment the rule set with more general, robust rules.
\end{itemize}
Both the \textit{guard\_model} and \textit{patch\_model} require stable, high-quality reasoning and are less likely to trigger refusal mechanisms. Therefore, we leveraged a powerful closed-source model with strong safety alignments, selecting \textbf{GPT-4o-mini} for these roles.

\paragraph{Validation Component} To prevent the augmented rules from overfitting, a validation step is performed using negative samples. The \textit{generate\_model} for creating these benign samples needs to be capable yet avoid refusal. We chose \textbf{DeepSeek-V3} for this role. For each positive (harmful) prompt, three corresponding negative (benign) samples were generated.

\subsection{Detailed Procedure}
The adversarial interaction proceeds as follows:
\begin{enumerate}
    \item The Red Team is given a set of harmful prompts (e.g., from HarmBench). For each prompt, it conducts up to 3 outer rounds of strategy refinement, with up to 10 inner attack attempts per round, to generate a set of effective jailbreak prompts that bypass the Blue Team's current guardrail ($Guard_{i-1}$).
    \item The Blue Team receives the set of successful jailbreak prompts ($\mathcal{A}_i$) from the Red Team. It uses the \textit{patch\_model} to analyze these prompts, identify common patterns, and generate a new, more generalized set of defensive rules.
    \item The Blue Team integrates these new rules into its existing rule set to form an updated guardrail, $Guard_i$.
    \item To prevent overfitting and ensure the new rules do not incorrectly block harmless queries, the Blue Team validates $Guard_i$ against a set of benign prompts ($\mathcal{D}_{benign, i}$).
    \item The process repeats, with the Red Team now facing the hardened guardrail $Guard_i$ in the next iteration ($i+1$). A detailed flowchart of this procedure is shown in Figure X.
\end{enumerate}

\section{Metric Definitions}
\label{sec:appendix_metrics}

Here we provide the formal definitions for the metrics used in our evaluation. Let $\mathcal{D}_{harm}$ be the set of harmful test samples ($N_{total} = |\mathcal{D}_{harm}|$). Let $\mathcal{A}_j$ be the set of effective attack prompts generated by the red team in iteration $j$. Let $\mathcal{D}_{benign, j}$ be the set of benign (negative) samples for validation in iteration $j$. Let $Guard_i$ be the blue team's guardrail system after iteration $i$.

\begin{itemize}
    \item \textbf{Defense Success Rate (DSR):} Our primary metric, measuring the proportion of harmful samples in round $i$ that are successfully intercepted by the guardrail from the previous state, $Guard_{i-1}$. It is the inverse of the Attack Success Rate (ASR), calculated as $1 - \text{ASR}_i$.
    \begin{equation}
    \small %
        \text{DSR}_i = 1 - \frac{1}{N_{total}} \sum_{q \in \mathcal{D}_{harm}} \text{IsSuccess}(p_{q,i}, G_{i-1})
    \end{equation}
    where $\text{IsSuccess}(p, G)$ returns 1 if prompt $p$ bypasses guardrail $G$.

    \item \textbf{Average Attack Turns (AAT):} The average number of inner attack attempts the Red Team requires for a successful jailbreak in round $i$.
    \begin{equation}
        \text{AAT}_i = \frac{\sum_{q \in \mathcal{S}_i} k_{q,i}}{|\mathcal{S}_i|}
    \end{equation}
    where $\mathcal{S}_i$ is the subset of successfully attacked samples and $k_{q,i}$ is the number of attempts for sample $q$.

    \item \textbf{Cross-Round Defense Efficacy (CRDE):} Formerly DSR, this metric measures the capability of the guardrail at iteration $i$ to defend against the set of effective attacks from a previous or current iteration $j$ (where $j \le i$). It tests for catastrophic forgetting.
\begin{equation}
\small
    \text{CRDE}_{i,j} = \frac{1}{|\mathcal{A}_j|} \sum_{p \in \mathcal{A}_j} \left(1 - \text{IsSuccess}(p, G_i)\right)
\end{equation}

    \item \textbf{False Positive Rate (FPR):} The proportion of harmless samples from $\mathcal{D}_{benign, j}$ that are incorrectly blocked by $Guard_i$.
    \begin{equation}
        \text{FPR}_{i,j} = \frac{\sum_{p_b \in \mathcal{D}_{benign, j}} \text{IsBlocked}(p_b, \text{Guard}_i)}{|\mathcal{D}_{benign, j}|} 
    \end{equation}
    where $\text{IsBlocked}(p_b, G)$ returns 1 if the benign prompt $p_b$ is incorrectly intercepted.
\end{itemize}

\section{Additional Results and Figures}
\label{sec:appendix_additional_results}

\begin{figure}[h!]
    \centering
    \includegraphics[width=1.0\linewidth]{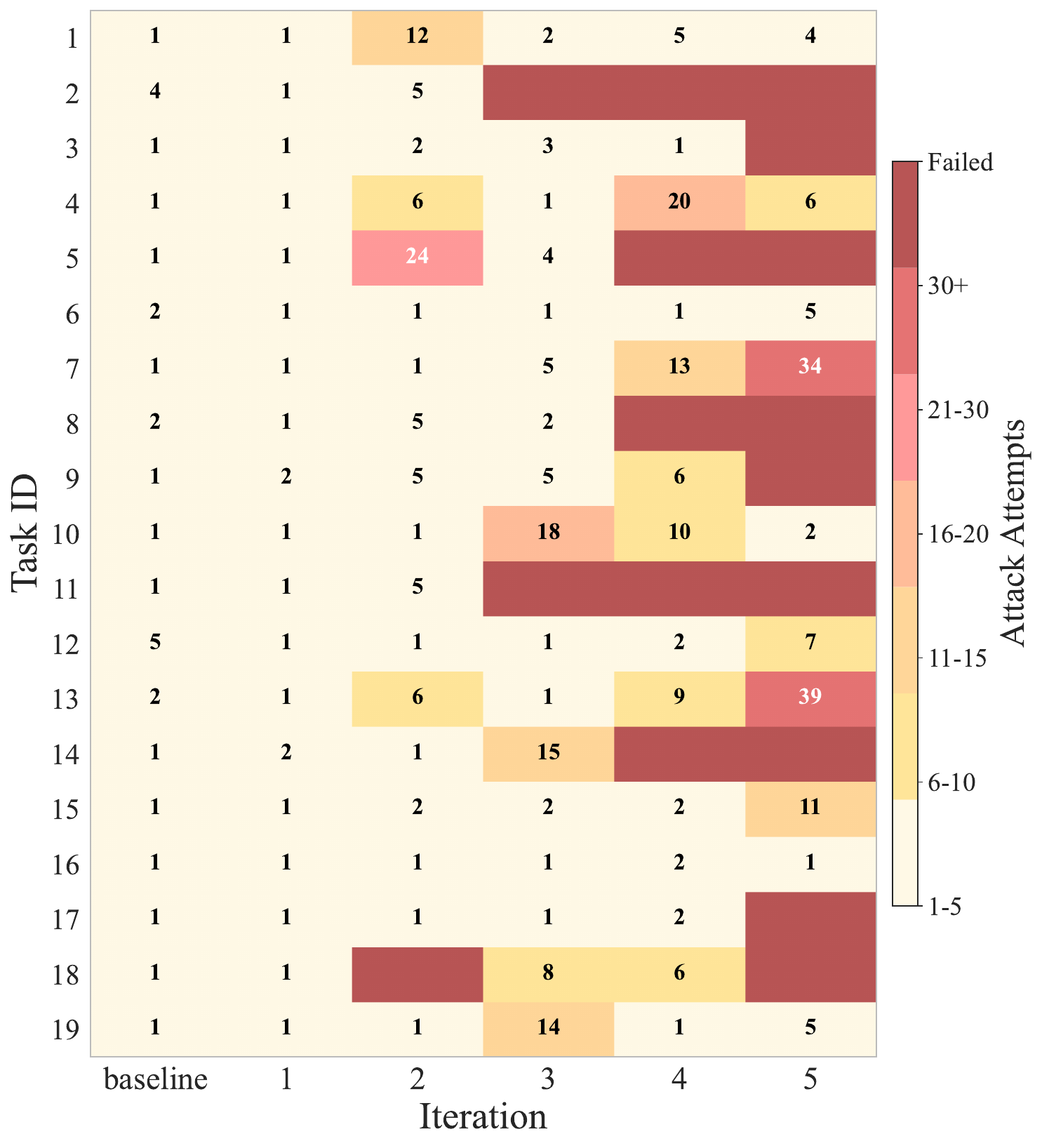}
    \vspace{-10pt}
    \caption{The Attack Time (number of attempts) of each task across all the adversarial rounds. A clear trend of increasing attack difficulty (color shifting from yellow to red/darker) is visible for most tasks in later iterations, complementing the aggregate AAT trend shown in Figure \ref{fig:dsr_aat}.}
    \label{fig:jailbreak_heatmap}
\end{figure}

This section provides supplementary figures referenced in the main body. Figure \ref{fig:jailbreak_heatmap} provides a task-level breakdown of the attack difficulty across the adversarial rounds.

\section{Cost Analysis}
\label{sec:cost}

To evaluate the operational viability of the RvB framework, we conduct a comprehensive analysis of computational costs in terms of token consumption. Contrary to the intuition that iterative adversarial loops increase overhead, our results demonstrate that the RvB paradigm is more resource-efficient than traditional cooperative Multi-Agent Systems (MAS).

\subsection{The Attacker vs. Defender in Cyber Security}

Table~\ref{tab:token_cost} details the token usage across the same experimental settings(Max\_epoch=5). The RvB framework consumes a total of 5,105,759 tokens, representing an \textbf{18.35\%} reduction in total computational cost compared to the Cooperative MAS (Pure Blue) baseline (6,253,192 tokens). 

We attribute this efficiency to the \textit{precision} of adversarial feedback. In a cooperative MAS environment, the defender often performs broad, indiscriminate reasoning to identify vulnerabilities, leading to high input token counts. Conversely, the Red Team in the RvB framework provides targeted, structured attack logs that serve as precise actionable contexts. This focused evidence prunes the Blue Team's search space, which is mathematically equivalent to reducing the epistemic uncertainty and entropy of the defense policy. Consequently, the Blue Team achieves higher remediation quality with fewer redundant reasoning cycles.

\begin{table}[ht]
\centering
\small %
\caption{Comparison of cumulative token consumption during the hardening process. The RvB framework significantly reduces input token overhead by leveraging targeted adversarial reports, which prevent redundant exploratory reasoning in the Blue Team.}
\label{tab:token_cost}
\begin{tabular}{lrrr}
\toprule
\textbf{Setting} & \textbf{Input} & \textbf{Output} & \textbf{Total} \\
\midrule
\begin{tabular}{@{}l@{}}Pure Red \\ (Offense)\end{tabular}           & 576,140    & 17,742   & 593,882   \\
\begin{tabular}{@{}l@{}}Pure Blue \\ (Cooperative MAS)\end{tabular}  & 5,943,099  & 310,093  & 6,253,192 \\
\textbf{RvB (Ours)}          & \textbf{4,800,430} & \textbf{305,329} & \textbf{5,105,759} \\
\bottomrule
\end{tabular}
\end{table}

\subsection{The Jailbreaker vs. Guardrail in Content Security}

In the content security experiment, we analyze the computational cost to demonstrate the framework's efficiency and accessibility. A key characteristic of this setup is its primary reliance on high-capability open-source models, allowing for local deployment. To provide a tangible measure of the resources consumed, we estimate the hypothetical cost of each round using publicly available reference API prices\footnote{Reference pricing data sourced from Third-party Platform}. The pricing for all models utilized in this experiment is detailed in Table~\ref{tab:content_api_costs}.

\begin{table}[ht]
\centering
\small
\caption{Reference API pricing for models used in the content security experiment. Prices are listed per million (M) tokens.}
\label{tab:content_api_costs}
\begin{tabular}{lrr}
\toprule
\textbf{Model} & \textbf{Input Prices} & \textbf{Output Prices} \\
\midrule
DeepSeek-V3 & \$0.2820 & \$1.1270 \\
Qwen3-32B & \$0.1410 & \$0.5640 \\
GPT-4o-mini & \$0.0970 & \$0.3870 \\
Qwen2-7B-Instruct & \$0.0000 & \$0.0000 \\
\bottomrule
\end{tabular}
\end{table}

Table~\ref{tab:content_token_cost} details the total token usage and the corresponding estimated costs for a static baseline and each iterative round of the RvB process. The costs are calculated by applying the reference prices from Table~\ref{tab:content_api_costs} to the precise input and output token consumption of each model involved in each stage.

\begin{table}[ht]
\centering
\small
\caption{Comparison of token consumption and estimated costs per round for the content security experiment.}
\label{tab:content_token_cost}
\begin{tabular}{lrr}
\toprule
\textbf{Round / Setting} & \textbf{Total Tokens} & \textbf{Estimated Cost (\$)} \\
\midrule
Baseline    & 4,040,889     & \$1.42 \\
\midrule
RvB Round 1                  & 2,828,789     & \$1.00 \\
RvB Round 2                  & 5,961,159     & \$2.10 \\
RvB Round 3                  & 10,490,319    & \$3.70 \\
RvB Round 4                  & 12,223,668    & \$4.31 \\
\bottomrule
\end{tabular}
\end{table}

The results clearly indicate the economic viability of our framework. Even the most computationally intensive iteration, Round 4, has a hypothetical cost of only \$4.31. This demonstrates that, even when benchmarked against commercial API rates, the per-round overhead of our iterative adversarial framework is well within an acceptable and modest range for academic research and practical application. This highlights the framework's accessibility, making advanced, co-evolutionary security hardening a financially feasible endeavor.

\end{document}